\pdfoutput=1
\RequirePackage[T1]{fontenc}
\documentclass[12pt]{article}

\usepackage[height=8.85in,width=6.45in]{geometry}

\usepackage[utf8]{inputenc}
\usepackage{amsmath}
\usepackage{amssymb}
\usepackage{mathtools}
\numberwithin{equation}{section}
\usepackage{slashed}
\usepackage{braket}
\usepackage[svgnames]{xcolor}
\usepackage[colorlinks,citecolor=DarkGreen,linkcolor=FireBrick,urlcolor=FireBrick,linktocpage,unicode]{hyperref}
\usepackage{cite}
\usepackage{graphicx}
\usepackage{tikz}
\tikzset{>=stealth}

\usepackage{times}
\usepackage{courier}
\usepackage{bm}
\usepackage{subfig}

\usepackage{xcolor}
\usepackage{mdframed}

\renewenvironment{figure}[1][]{
  \begin{originalfigure}[#1]
    \begin{mdframed}[linecolor=black!0,backgroundcolor=black!1]
}{
    \end{mdframed}
  \end{originalfigure}
}

\def\Nequals#1{$\mathcal{N}{=}#1$}
\def\bZ{\mathbb{Z}}
\def\MF{\mathrm{MF}}
\def\TMF{\mathrm{TMF}}
\def\Tmf{\mathrm{Tmf}}
\def\tmf{\mathrm{tmf}}
\def\tr{\mathop{\mathrm{tr}}}

\begin{document}

\begin{titlepage}

\begin{flushright}
\end{flushright}

\vskip 3cm

\begin{center}

{\Large \bfseries Topological modular forms and\\[1em]
 the absence of a heterotic global anomaly}

\vskip 1cm
 Yuji Tachikawa
\vskip 1cm

\begin{tabular}{ll}
  &Kavli Institute for the Physics and Mathematics of the Universe, \\
 & University of Tokyo,  Kashiwa, Chiba 277-8583, Japan
\end{tabular}

\vskip 1cm

\end{center}

\noindent

Spacetime theories obtained from perturbative string theory constructions are automatically free of perturbative anomalies,
but it is not settled whether they are always free of global anomalies.
Here we discuss a possible $\bZ_{24}$-valued pure gravitational anomaly
of heterotic compactifications down to two spacetime dimensions,
and point out that it can be shown to vanish using the theory of topological modular forms,
assuming the validity of the Stolz-Teichner conjecture.

\end{titlepage}

\setcounter{tocdepth}{3}
\tableofcontents

\section{Introduction and summary}
\label{sec:introduction}

In the last several years, we have seen a significant progress in the study of anomalies in general.
Although the nature of perturbative anomalies were extensively analyzed in the '80s,
there was no systematic analysis of global anomalies, 
which were studied only in a case-by-case basis.
In part stimulated by the developments in theoretical condensed matter physics, 
we now have a unifying framework to understand both perturbative and global anomalies \cite{Freed:2016rqq}.
This short paper records a small step in the application of the new methods to the question of anomalies in string theory.

For the purpose of this paper, we regard  perturbative string theory as an elaborate machinery which produces a $d$-dimensional spacetime theory from a two-dimensional conformal field theory (2d CFT) with appropriate properties.
The 2d CFT in question can arise, but not necessarily, as the worldsheet sigma model of the string in an internal manifold of dimension $10-d$,
in which case  the analysis of  perturbative and global anomalies on the worldsheet is required.
We take the position that this has been taken care of if necessary
so that we have a consistent 2d CFT at hand as an input.
We can then concentrate on the question of anomalies of the spacetime theory.

More specifically, we consider the case of perturbative heterotic string theories compactified down to $d$ dimensions,
for which the input is 
\begin{equation}
\text{an \Nequals{(0,1)} supersymmetric 2d CFT with  
$(c_L,c_R)=(16+(10-d),\frac32(10-d))$}
\end{equation}
and the output is \begin{equation}
\text{a spacetime theory in $d$ dimensions}.
\end{equation}
The resulting spacetime theory is famously free of perturbative anomalies; 
for $d{=}10$ it is the foundational result of Green and Schwarz \cite{Green:1984sg},
which was later generalized to arbitrary $d$ by Schellekens, Warner and collaborators in \cite{Schellekens:1986xh,Lerche:1987qk,Lerche:1988np}.
The absence of  global anomalies was studied in \cite{Witten:1985xe,Witten:1985bt} for $d=10$,
but the case in general dimensions has not been systematically explored to the authors' knowledge.\footnote{%
Even the question whether the 4d heterotic compactifications are free of Witten's $SU(2)$ anomaly is wide open;
see \cite{Enoki:2020wel} for a recent discussion in the context of 4d \Nequals2 compactifications.
}

Let us first introduce the particular anomaly we are going to study in this note,
and then put it in a more general perspective. 
We consider heterotic compactifications to two dimensions.
It can have a chiral fermionic spectrum, say with $n$ left-moving Majorana-Weyl fermions, which can contribute to the anomaly polynomial of the form \begin{equation}
-n(p_1/48),\label{f}
\end{equation}
where $p_1$ is the first Pontryagin class of the spacetime tangent bundle.
We can try to cancel this anomaly by the Green-Schwarz mechanism, by including the two-dimensional coupling 
\begin{equation}
2\pi iN\int B,\label{B}
\end{equation}
where the $B$-field is normalized to have periodicity one.
In heterotic strings, we know that the field strength $H$ of $B$ satisfies \begin{equation}
dH=p_1/2.\label{H}
\end{equation}
The normalization of the right hand side can be explained in multiple ways,
some of which will be explained later.
Comparing \eqref{f} and \eqref{H}, we see that the fermionic anomaly \eqref{f} can be canceled by setting $N=n/24$.
The invariance of the coupling \eqref{B} under the large gauge transformation of the $B$ field  requires that $N$ is an integer,
implying that there can be a $\bZ_{24}$-valued global anomaly in this system,
unless $n$ is divisible by $24$.

Let us put it in a broader context.
To discuss global anomalies from the modern point of view, it is imperative to specify the spacetime structure.
We assume that the spacetime is oriented and equipped with a spin structure.
In addition, we assume that a 3-form field strength $H$ satisfying the relation \eqref{H} is specified.
This combination of geometric data is known as a string structure in the mathematical literature, 
which can also be motivated purely within algebraic topology.\footnote{
Namely, we consider  trivializing a given orthogonal bundle.
The first obstruction is  $w_1\in H^1(BO,\bZ_2)$ which is trivialized by the orientation,
the second obstruction is $w_2\in H^2(BSO,\bZ_2)$ which is trivialized by the spin structure,
and the next obstruction  is $\lambda\in H^4(BSpin,\bZ)=\bZ$
which satisfies $2\lambda=p_1$ and is trivialized by the string structure.}
The possible global anomaly in $d$ dimensions is then governed by the torsion part of $\Omega^\text{string}_{d+1}$,
the string bordism group in $d+1$ dimensions, which is given as follows \cite{Giambalvo1971}:
\begin{equation}
\begin{array}{c|ccccccccccccccccccccc}
D=d+1& 0 & 1 & 2 & 3 & 4 & 5 & 6 & 7 & 8 & 9 & 10 & 11 & 12  & 13 
\\
\hline
\Omega^\text{string}_D &
\bZ &
\bZ_2 &
\bZ_2 &
\bZ_{24} &
0 &
0 &
\bZ_2 & 
0 &
\bZ\oplus \bZ_2 &
(\bZ_2)^2 &
\bZ_6&
0 &
\bZ &
\bZ_3 
\end{array}.
\end{equation}
What we saw above is the $\bZ_{24}$-valued global anomaly in the case $d=2$.

As we will see below in more detail,
the absence of this global anomaly boils down to the question 
whether the net number $b$ of chiral Majorana-Weyl fermions in two-dimensional heterotic compactifications
is always divisible by 24.
In terms of the property of 2d \Nequals{(0,1)} CFT $T$ with $(c_L,c_R)=(24,12)$ on the worldsheet, 
this number $b$ appears in the expression of its elliptic genus  as the constant term in its $q$-expansion: \begin{equation}
Z_\text{ell}(T;q) = a q^{-1} + b + O(q). \label{eq:exp}
\end{equation}
The question is then whether this coefficient $b$ is divisible by 24 for any such CFTs.

The author does not know of any field theoretical technique to study this divisibility.
The only approach that he is aware of is to utilize the theory of topological modular forms,
which goes as follows; more details will be provided later.

Hopkins and collaborators have constructed  a generalized homology theory  $\TMF_\bullet$ known as topological modular forms \cite{Hopkins1994,Hopkins2002,TMFBook}.
What makes $\TMF$ useful for us is the conjecture of Stolz and Teichner \cite{StolzTeichner1,StolzTeichner2},
which says that the group $\TMF_\nu$ is the group of deformation classes of 
2d \Nequals{(0,1)} supersymmetric field theories with gravitational anomaly\footnote{%
It is a somewhat confusing point in this paper that 
the worldsheet and the spacetime are both two dimensional.
We denote the anomaly of the spacetime as $n(p_1/48)$ and 
that of the internal worldsheet CFT as $\nu(p_1/48)$.
} $\nu(p_1/48)$.
Then a 2d \Nequals{(0,1)} CFT $T$ determines a class $[T]\in \TMF_{-2(c_L-c_R)}$.

Now, $\TMF_\bullet$ has a natural homomorphism $\varphi_W$, called the Witten genus, to the ring of integral modular forms $\MF_{\bullet/2}$ with the modular discriminant $\Delta$ inverted:\begin{equation}
\varphi_W: \TMF_\bullet \to \MF[\Delta^{-1}]_{\bullet/2}.
\end{equation}  
Under the Stolz-Teichner conjecture, the Witten genus is equal to the elliptic genus as defined by physicists, multiplied by $\eta(q)^\nu$: \begin{equation}
\varphi_W([T])=\eta(q)^{\nu} Z_\text{ell}(T;q).
\end{equation}
The fact that the image of $\varphi_W$ is completely determined \cite[Proposition 4.6]{Hopkins2002}
then allows us to conclude that $b$ in \eqref{eq:exp} is a multiple of 24,
showing that the heterotic compactifications to two dimensions are free of the $\bZ_{24}$-valued 
pure gravitational global anomaly.

The rest of  the paper is organized as follows.
In Sec.~\ref{sec:anomaly}, we will describe how
a single Majorana-Weyl spinor in two dimensions has an anomaly $1$ mod 24,
and recast the condition of the absence of the anomaly in terms of the property of the elliptic genus $Z_\text{ell}$ of the internal  worldsheet CFT.
In Sec.~\ref{sec:tmf}, we very briefly review basic properties of the topological modular forms,
their relation to integral modular forms, and the Stolz-Teichner conjecture.
We will see that the coefficient $b$ is indeed a multiple of $24$.

Before proceeding, we would like to mention that in this short note we are only going to scratch the surface of the relationship between $\TMF$ and the anomaly of heterotic compactifications.
A more systematic analysis will be presented elsewhere.

\section{A heterotic global anomaly}
\label{sec:anomaly}
\subsection{The $\bZ_{24}$ anomaly}
We start our discussion with the gravitational anomaly of $n$ left-moving Majorana-Weyl fermions in two dimensions.
Its anomaly polynomial is given by $-np_1/48$.
This has the following interpretation: the Majorana-Weyl fermions live on the boundary of a fermionic invertible phase in three dimensions, whose partition function on a closed spin manifold $M_3$ is given by \begin{equation}
 e^{2\pi i n \eta(M_3)  }= e^{-2\pi i n \int_{W_4} p_1/48 }.
\end{equation}
Here, $\eta(M_3)$ is the normalized eta invariant appropriate for a single 2d Majorana-Weyl spinor,
and $W_4$ is a spin 4-manifold such that $\partial W_4=M_3$.
The existence of $W_4$ is guaranteed thanks to $\Omega^\text{spin}_3=0$,
and $Z_{M_3}$ does not depend on the choice of $W_4$ 
because $p_1/48$ integrates to an integer on any spin 4-manifold.

We can try to cancel this anomaly via the Green-Schwarz mechanism. 
Namely, we introduce a $B$-field whose 3-form field strength $H$ satisfies $dH=p_1/2$.\footnote{%
Here it is important that we cannot make the denominator larger.
In a more proper formulation of the $B$-field, we need to require that the right hand side is well-defined also as an integral cohomology class;
there is an integral class $\lambda$ generating $H^4(BSpin,\bZ)\simeq\bZ$ which satisfies $2\lambda =p_1$,
and we use the pull-back of $\lambda$ via the classifying map to $BSpin$ to formulate this relation.
}
We then consider a coupling $2\pi i N\int_{\Sigma_2} B$ in two dimensions.
This corresponds to the inclusion of the contribution \begin{equation}
 e^{2\pi i N \int_{M_3} H} = e^{2\pi i N \int_{W_4}p_1/2} 
\end{equation} in the bulk action,
which allows us to cancel the fermionic anomaly if $n=24N$, i.e.~when $n$ is divisible by $24$.\footnote{%
We note that this is also how the heterotic worldsheet theory is anomaly free.
Indeed, in the fermionic formulation in the light-cone gauge, there are 32 left-moving Majorana-Weyl fermions for the current algebra,
and there are eight right-moving superpartners of the eight spatial coordinates. 
In total the anomaly is $(32-8)(-p_1/48)=-p_1/2$, which is canceled by the anomalous variation of the $B$ field coupling with $N=1$.
}

We can try to include the $B$-field coupling with $N=n/24$ even when $N$ is fractional.
In this case, there is a $\bZ_{24}$-valued global anomaly under the large gauge transformation of the $B$-field.\footnote{%
A six-dimensional analogue of this construction was discussed in \cite{Lee:2020ewl}.
}
Taking $n=1$, the invertible phase for the string structure in three dimensions is then given by \begin{equation}
e^{2\pi i (\eta(M_3) +\int_{M_3} H/24)}.
\end{equation}
The expression above evaluates to $e^{-2\pi i/24}$ on $S^3$ with $\int_{S^3}H=1$,
which is known to generate $\Omega^\text{string}_3=\bZ_{24}$.

\subsection{Heterotic compactifications to two dimensions}
Let us now point out how this $\bZ_{24}$ anomaly could arise in the context of heterotic string compactifications down to two dimensions.
Except for the analysis of the possible anomaly,
the content of this subsection is totally standard and has been explored in detail in the past, 
see e.g.~\cite{Vafa:1995fj,Sen:1996na,Bergman:2003ym,Font:2004et,Paquette:2016xoo,Apruzzi:2016iac,Florakis:2017zep,Melnikov:2017yvz}.

We take an \Nequals{(0,1)}-supersymmetric CFT $T$ with $(c_L,c_R)=(16+8,\frac32 \cdot 8)$
as the internal worldsheet degrees of freedom.
To obtain spacetime massless fermions,
the standard rule of the string perturbation theory tells us that the world-sheet right-movers are in the R-sector vacuum,
and the world-sheet left-movers can have either $L_0=0$ or $1$.
More precisely, \begin{itemize}
\item states with $L_0=0$ give rise to $\psi^\pm_\mu$ and $\lambda^\mp$, depending on $(-1)^{F_R}=\pm1$, and 
\item states with $L_0=1$ give rise to $\psi^\pm$, again depending on  $(-1)^{F_R}=\pm1$.
\end{itemize}
Here $\psi_\mu$ are gravitinos while $\lambda$ and $\psi$ are ordinary fermions;
the superscripts $\pm$ and $\mp$ specify whether the modes are left-moving or right-moving,
and $(-1)^{F_R}$ is the right-moving fermion number.

We now recall that the number of states weighted by $(-1)^{F_R}$ is encoded in 
the elliptic genus of the internal CFT $T$, which has the following form: \begin{equation}
Z_\text{ell}(T;q) = \tr (-1)^{F_R} q^{L_0 - \frac{c_L}{24}}  = a q^{-1} + b  + \cdots
= a J(q) + b,
\end{equation}
where $a,b\in\bZ$.
In the last equality we used the fact that the elliptic genus is modular;
$J(q)$ is the modular $j$ function normalized to have the leading term $q^{-1}$ and no constant term.

The anomaly can then be  computed, using the fact that the anomaly polynomial of $\psi^\pm_\mu$ and $\psi^\pm$ are $\pm 23 p_1/48$ and $\mp p_1/48$, respectively \cite{AlvarezGaume:1983ig}.
We find that the total anomaly polynomial is given by \begin{equation}
(-24a + b) (-p_1/48),
\end{equation} which requires the spacetime coupling $2\pi i N\int B$ with \begin{equation}
N= -a + {b}/{24}.
\label{N}
\end{equation}
As in the original case of Green and Schwarz \cite{Green:1984sg}
and also in higher dimensions \cite{Schellekens:1986xh,Lerche:1987qk,Lerche:1988np}, 
the $B$-field coupling with this required strength is automatically generated in the string 1-loop perturbation theory \cite{Vafa:1995fj,Sen:1996na}:
the integral of the $B$-field vertex operator reduces to the expression \begin{equation}
N= \frac{1}{8\pi}\int_{\mathcal{M}}  Z_\text{ell}(T;q) d\mu,
\end{equation}
where $\mathcal{M}$ is the fundamental region of $SL(2,\bZ)$ and $d\mu=dxdy/y^2$ for $\tau=x+iy$ is the standard measure on it.
This expression can be evaluated to give \eqref{N}.

When $N$ is an integer,  
we can include appropriately-oriented $|N|$ space-filling heterotic strings to cancel the $B$-field tadpole.
When $N$ is fractional, 
it presents a genuine global anomaly of a heterotic compactification down to two dimensions.
As mentioned, many  heterotic compactifications down to two dimensions have been studied,
but the coefficient $b$ was divisible by $24$ in all known examples.
The question then is whether this property holds for arbitrary \Nequals{(0,1)}-supersymmetric CFT with $(c_L,c_R)=(24,12)$.

\section{Its absence via topological modular forms}
\label{sec:tmf}
\subsection{$\tmf$ and the Witten genus}
Thus far, we reinterpreted the question of the global anomaly of heterotic compactifications to two dimensions into the divisibility by 24 of the constant term $b$ of the $q$ expansion of the elliptic genus of the corresponding worldsheet CFT.
The remaining task is to show that $b$ is indeed divisible by $24$,
using the theory of topological modular forms.

Topological modular forms come in three closely-related  variants, $\TMF$, $\Tmf$ and $\tmf$.
They are generalized homology theories constructed by Hopkins and collaborators 
by combining algebraic geometry and algebraic topology \cite{Hopkins1994,Hopkins2002,TMFBook}.
Here we only cite the properties we will need.
First, $\tmf$ is connective, i.e.~$\tmf_{\bullet<0}=0$,
and a $\nu$-dimensional manifold $M$ with string structure determines a class $[M]\in \tmf_\nu$.
In more detail, the class $[M]$ only depends on the string bordism class in $\Omega^\text{string}_\nu$,
and therefore there is a homomorphism $\Omega^\text{string}_\bullet \to \tmf_\bullet$.
This homomorphism is known to be surjective.

In addition, there is a homomorphism \begin{equation}
\varphi_W: \tmf_{\bullet} \to \MF_{\bullet/2}
\end{equation}
where  $\MF=\bZ[c_4,c_6,\Delta]/(c_4^3-c_6^2-1728\Delta)$ is the ring of integral modular forms,
with  \begin{equation}
c_4 = 1+240 q + \cdots,\qquad
c_6 = 1-504 q - \cdots
\end{equation}being the standard normalized Eisenstein series and \begin{equation}
\Delta=q - 24 q^2 + \cdots 
\end{equation}  the modular discriminant satisfying $1728\Delta=c_4^3-c_6^2$.
Here we put $c_4$, $c_6$ and $\Delta$ in degree $4$, $6$ and $12$ of $\MF_\bullet$.
This homomorphism $\varphi_W$ is known as the Witten genus.

The proposition 4.6 of \cite{Hopkins2002} describes the image of $\varphi_W$. 
Namely,  the image of $\varphi_W$ has a $\bZ$-basis given by \begin{equation}
a_{i,j,k} c_4^i c_6^j \Delta^k, \qquad \text{$i\ge 0$; $j=0,1$; $k\ge 0$}
\end{equation}where \begin{equation}
a_{i,j,k} = \begin{cases}
24/\gcd(24,k) & \text{if $i=j=0$},\\
2 & \text{if $j=1$},\\
1 & \text{otherwise}.
\end{cases}
\end{equation}
The kernel of $\varphi_W$ consists of torsion elements of $\tmf$.
In particular, the kernel is known to be absent when $k$ is a multiple of $24$.

It is known that $\varphi_W$ applied to $[M]$ satisfies the following equation \begin{equation}
\varphi_W([M])= \eta(q)^{\nu} q^{-\nu/24}\int_M \hat{A}(R) \tr\prod_{\ell=1}^\infty (1-q^\ell  e^{iR/2\pi})^{-1}
\label{sigma-model-witten-genus}
\end{equation}  where $R$ is the curvature 2-form.
Let us recall its physical significance \cite{Witten:1986bf,Zagier1986,Witten:1987cg}.
We consider the \Nequals{(0,1)} sigma model with the target space $M$.
The right-moving fermions have sigma-model anomalies in general \cite{Moore:1984dc},
which can be canceled by introducing a coupling to the $B$ field, if its field strength $H$ can be arranged to solve $dH=p_1/2$.
More precisely, this equality has to hold at the level of integral cohomology \cite{Witten:1985mj},
and the data of such $H$ together with the spin structure of $M$ comprise the string structure of $M$.
Summarizing, given an $n$-dimensional manifold $M$ with string structure,
we can consider the \Nequals{(0,1)} sigma model $\sigma_M$ whose target space is $M$.
Its elliptic genus $Z_\text{ell}(\sigma_M;q)$ can be computed in the large volume limit, 
and is given by \eqref{sigma-model-witten-genus} without the prefactor $\eta(q)^\nu$, 
so that we have the relation \begin{equation}
\varphi_W([M]) = \eta(q)^\nu Z_\text{ell}(\sigma_M;q).
\end{equation}

\subsection{$\TMF$ and a tentative solution to our question}

The discussions above strongly suggest that  topological modular forms and 2d \Nequals{(0,1)} theories should be related.
Indeed, Stolz and Teichner \cite{StolzTeichner1,StolzTeichner2} conjectured that another version of topological modular forms, $\TMF$,
classifies 2d \Nequals{(0,1)} theories in the following manner. 
We first note that $\Delta^{24}$ is in the image of $\varphi_W$ applied to $\tmf_{576}$ according to the theorem above.
Furthermore, the kernel of $\varphi_W$ is trivial at that degree.
Therefore there is a unique element which maps to $\Delta^{24}$ under $\varphi_W$, which we also denote by $\Delta^{24}$.
Then $\TMF$ is obtained by inverting $\Delta^{24}$: \begin{equation}
\TMF_\bullet=  \tmf[\Delta^{-24}]_\bullet,
\end{equation}
which is now $576$-periodic.
The conjecture of Stolz and Teichner is then \begin{equation}
\TMF_\nu = \left\{
\begin{array}{c}
\text{2d \Nequals{(0,1)}-supersymmetric QFTs}\\
\text{whose gravitational anomaly is $\nu(p_1/48)$}
\end{array}\right\} / \sim,
\end{equation} where two theories $T$ and $T'$ are considered equivalent $T\sim T'$ when they can be connected by continuous (not necessarily marginal) deformations,
and with the convention on $\nu$ that a single chiral multiplet has $\nu=+1$ instead of $\nu=-1$.
This conjecture has  received some attention on the theoretical physics side recently
\cite{Gaiotto:2018ypj,Gukov:2018iiq,Gaiotto:2019asa,Gaiotto:2019gef,Johnson-Freyd:2020itv},
in which various pieces of evidence toward the validity of the conjecture can be found.
We extend the Witten genus to the homomorphism \begin{equation}
\varphi_W : \TMF_\bullet \to \MF[\Delta^{-1}]_{\bullet/2}.
\end{equation}
We assume that $\varphi_W$ applied to the class $[T]\in \TMF$ satisfies the following relationship as before: \begin{equation}
\varphi_W([T])=\eta(q)^\nu Z_\text{ell}(T;q).
\end{equation}

After these preparations, the desired divisibility of $b$ by $24$ is an immediate corollary. 
Let $T$ be the internal worldsheet CFT to be used in a  heterotic compactification down to two dimensions.
It has $(c_L,c_R)=(16+8,\frac32 8)$,
which corresponds to the gravitational anomaly $\nu=2(c_R-c_L)=-24$.
The Witten genus is then \begin{equation}
\varphi_W([T]) = \eta(q)^{-24} (a J(q) + b ) = a c_4^3\Delta^{-2} + (-744a+b)\Delta^{-1}.
\end{equation}
Now we invoke the proposition above, which says that $\varphi_W(\TMF_{-24})$ has an integral basis
consisting of $c_4^{3i}\Delta^{-(i+1)}$ for $i\ge1$ and $24\Delta^{-1}$.
This implies that $b$ is a multiple of $24$.

\section*{Acknowledgements}
The author would  like to thank the authors of \cite{Enoki:2020wel} for a tea-time chat reminding him that the question of the global anomaly of generic heterotic compactifications is wide open,
Kantaro Ohmori for stimulating discussions.
He also would like to thank Mohab Abou Zeid, I\~naki Garc\'\i a-Etxebarria, Abhiram Kidambi, Yasunori Lee and Taizan Watari for helpful comments on earlier drafts.
He is partially supported  by JSPS KAKENHI Grant-in-Aid (Wakate-A), No.17H04837 
and JSPS KAKENHI Grant-in-Aid (Kiban-S), No.16H06335,
and also by WPI Initiative, MEXT, Japan at IPMU, the University of Tokyo.

\def\arxivfont{\rm}
\bibliographystyle{ytamsalpha}
\baselineskip=.93\baselineskip
\let\originalthebibliography\thebibliography
\renewcommand\thebibliography[1]{
  \originalthebibliography{#1}
  \setlength{\itemsep}{0pt plus 0.3ex}
}
\bibliography{ref}

\end{document}